\documentclass[useAMS,usenatbib,referee]{mn2e}
\usepackage{epsfig}
\title[Gamma-ray spectrum of Vela-like pulsars]
{Outer-magnetospheric model for Vela-like pulsars: Formation of sub-GeV spectrum}
\author[J.~Takata, S.~Shibata \& K.~Hirotani]
  {J.~Takata,$^1$\thanks{takata@ksirius.kj.yamagata-u.ac.jp}
  S.~Shibata,$^2$ K.~Hirotani,$^3$ \\
  $^1$Graduate School of Science and Engineering, Yamagata University, Yamagata
990-8560, Japan\\
  $^2$Department of Physics, Yamagata University, Yamagata 990-8560, Japan \\
  $^3$Max-Planck-Institut fuer Kernphysik, Postfach 103980 D-69029 Heidelberg, Germany }
\date{Released 2003 Xxxxx XX}

\pagerange{\pageref{firstpage}--\pageref{lastpage}} \pubyear{2003}

\def\LaTeX{L\kern-.36em\raise.3ex\hbox{a}\kern-.15em
    T\kern-.1667em\lower.7ex\hbox{E}\kern-.125emX}

\begin{document}

\label{firstpage}

\maketitle

\begin{abstract}
We investigate the $\gamma$-ray emission from a outer-gap accelerator, 
which is located in  
the outer-magnetosphere of a pulsar. The charge depletion from 
the Goldreich-Julian 
density causes a large electric field along the magnetic field lines. 
The electric field accelerates electrons and positrons to high energies. 
We solve the formation of the electric field self-consistently with 
 the radiation and the pair-creation processes in 
one-dimensional geometry along the magnetic field lines, 
and calculate curvature spectrum to compare with the 
observations. 

We find that because the particles escape from the accelerating region 
with large Lorentz factors as $10^{7.5}$, curvature radiation from the outside 
of the gap, in which there is no electric field, still has a important 
contribution to the spectrum. Including the outside emission, 
the model spectra are improved between 
100 MeV and 1 GeV for the Vela-type pulsars. 
Since the field line
 curvature close to  the light cylinder affects the spectrum 
in several hundred MeV, 
there is a possibility to diagnose the field structure by using 
the spectrum in these bands. 
The difference found in the $\gamma$-ray spectra of the twin 
pulsars, PSR~B0833-45 (Vela) and B1706-44, is deduced from the dimensionless 
electric current running through the gap. The extension to the 
three-dimensional modeling with self-consistent 
electrodynamics is necessary to understand 
the difference in the observed flux between the twin pulsars.   
\end{abstract}

\section{Introduction}
\label{intro}
The {\it Compton Gamma Ray Observatory} 
had detected seven $\gamma$-ray pulsars
(Thompson et al. 1999).
The observed light curves and energy spectra tell about the energy 
of accelerated particles and the geometry of $\gamma$-ray emission regions 
in the pulsar magnetosphere. These observations provide constraints 
on proposed radiation models.  
The next-generation $\gamma$-ray space telescopes and ground based 
\v Cerenkov telescopes are expected to provide further constraints 
on  the models.  
 
A pulsation in  $\gamma$-rays implies that the acceleration of particles
 and the subsequent radiation occur within the light cylinder,
 of which axial distance  is given by 
$\varpi_{lc}=c/\Omega$,
 where $\Omega$ is angular frequency of the star, and $c$ the  
speed of light.   

The charge depletion from  Goldreich-Julian 
charge density causes a large electric field $E_{||}$ 
along the magnetic field lines. The Goldreich-Julian  density is given by 
(Goldreich \& Julian 1969, here after GJ)
\begin{equation}
\rho_{GJ}=-\frac{\Omega B_z}{2\pi c}\frac{1}{[1-(\Omega\varpi/c)^2]},
\label{GJ}
\end{equation}
where $B_z$ is the component of the magnetic field along the rotation axis, 
and $\varpi$ the axial distance.  A potential drop in such a charge 
depletion region (gap) is only a small 
fraction of the available electromotive force exerted on the spinning neutron 
star surface. The $\gamma$-ray emissions in the pulsar magnetosphere 
 have been argued with polar-cap model 
(e.g. Sturrock 1971; Ruderman \& Sutherland 1975) and outer-gap model 
(e.g. Cheng, Ho \& Ruderman 1986a, b); 
each acceleration region is located near 
stellar surface above the magnetic poles and in the outer-magnetosphere 
around the null charge surface ($B_z=0$) above the last-open field line, 
respectively. 
      
After Cheng et al. (1986a,b), the outer-gap model has 
been studied in two different pictures.
 In the first picture, the observed light curves have been 
studied in an effort to construct the thee-dimensional models for the outer-gap
 (Romani \& Yadigaroglu 1995; Romani 1996; Cheng, Ruderman \& Zhang 2000). 
The peaks in the light curves are interpreted as an effect 
of aberration and time delay for emitted photons, thereby the outer-gap 
model has been successful in explaining the double peaks 
in one period and phase separation of the pulses in different energies from 
the radio to $\gamma$-ray.  The observed phase resolved 
spectra have also been calculated with the models, in which pair cascade 
process has been
solved with assumed field-aligned electric field $E_{||}$. 
The predicted phase resolved spectrum depends on the assumption of the strength
 and distribution of $E_{||}$. Thus, the models are not satisfactory 
in the sense of electrodynamics.
 
The other approach has focused on electrodynamics in the outer-gap.
 Cheng et al. (1986a) discussed the electric structure in the vacuum gap. 
They gave  a solution for the Poisson equation with rectilinear geometry.  
Hirotani \& Shibata (1999, hereafter HS) proposed a non-vacuum 
electrodynamical model for the outer-gap.  Although they worked 
on only one dimension along the last-open field line, 
they solved the accelerating electric field self-consistently with curvature
 radiation and pair-creation processes.
Although the calculated $\gamma$-ray spectra were consistent 
with the  EGRET  observation in the GeV band, 
there were much harder than what is observed below 1 GeV.

In this paper, to improve the HS model, we take account of the curvature 
radiation from the outside of the gap.  Even though the electric field is 
screened out in this region, particles still possess sufficient energy and 
hence contribute to the $\gamma$-ray luminosity  between 
100 MeV and 1 GeV. Quite recently, Hirotani, Harding \& Shibata 
(2003) presented a model in which particle motion is precisely solved under 
the electric force and radiation reaction force 
(see also Hirotani 2003 for a review). 
Although they also took account 
of the radiation from the outside of the gap, 
we treat this effect in some detail by considering non-dipole field 
near the light cylinder because the field line are expected to deviate from 
dipole configuration by the influence of the pulsar wind or the electric 
current. Then, we show that the position of low energy cut-off 
in the curvature spectrum depends on the curvature radius of the field lines 
near the light cylinder. In applying our model, we consider the twin pulsars, 
PSR~B0833-45 (Vela) and B1706-44.
 It is remarkable that these  pulsars show distinctive $\gamma$-ray spectra
 even though they have very 
similar magnetospheric parameters as well as surface 
temperatures (Gotthelf, Halpern \& Dodson 2002).  Thus 
this pair of pulsars gives unique opportunity to examine the model 
capability.  We shall show that the difference of spectral 
peak energy  between the two, about 3 GeV for Vela and about 1 GeV for 
B1706-44, is understood well with the present model. 
On the other hand, we elucidate 
limitations of one-dimensional approximation.

In \S\ref{model} we describe the basic equations for the electrodynamics 
in the gap. In \S\ref{outrad} we discuss the 
curvature radiation from the outside of the gap. In \S\ref{appli} we apply
 our model to the Vela pulsar and PSR~B1706-44. The necessity for the 
three-dimensional electrodynamical model is discussed in \S\ref{dicus}.  
 
\section{One dimensional outer-gap model}
\label{model}
In this section, we present the basic equations to describe the 
electrodynamics in the gap, after HS. We give the Poisson equation 
for the acceleration field (\S\S\ref{poission}), 
the continuity equations for particles and $\gamma$-rays 
(\S\S\ref{cont}). We impose the boundary conditions in \S\S\ref{bound} and 
finally describe the $X$-ray field in \S\S\ref{Xray}
\subsection{Poisson equation}
\label{poission}
If the magnetosphere is corotating with the star, an electric field, 
\begin{equation}
\bmath{E}=-\frac{(\bmath{\Omega}\times\bmath{r})\times\bmath{B}}{c}
\equiv-\nabla\Phi_{co},
\end{equation}
which is perpendicular to the magnetic field $\bmath{B}$, is exerted at 
position $\bmath{r}$. The corotational potential $\Phi_{co}$ satisfies the 
Poisson equation
\begin{equation}
\nabla^2\Phi_{co}=-4\pi\rho_{GJ}.
\label{poisson1}
\end{equation}
 Meanwhile in the gap, since the real 
charge density $\rho$ deviates from $\rho_{GJ}$, the electrostatic 
potential $\Phi$ differs from $\Phi_{co}$. The difference is extracted as the 
{\it non-corotational} potential,  $\Phi_{nco}=\Phi-\Phi_{co}$. 
The Poisson equation, $\nabla^2\Phi=- 4\pi\rho$, yields  
\begin{equation}
\nabla^2\Phi_{nco}=-4\pi\rho_{eff},
\label{poisson2}
\end{equation}
where $\rho_{eff}=\rho-\rho_{GJ}$ is the effective charge density which 
produces the 
field-aligned electric field, $E_{||}=-\bmath{B}\cdot\nabla\Phi_{nco}
/|\bmath{B}|$. 

If the gap width $W_{||}$ along the magnetic field lines is much shorter than 
the curvature radius ($\sim\varpi_{lc}$) of the field lines, 
we may assume that the field lines are straight in the gap. 
In \S\ref{appli}, we shall see that $W_{||}$ is  about 10 per cent of $\varpi_{lc}$. 
By assuming that the gap width is much smaller than the trans-field thickness 
of the gap, we rewrite equation (\ref{poisson2}) as
\begin{eqnarray}
\frac{d^2\Phi_{nco}(s)}{ds^2}
&=&-4\pi\left[\rho(s)-\rho_{GJ}(s)\right] \nonumber \\
&=&-4\pi e\left[N_{+}(s)-N_{-}(s)-\frac{\rho_{GJ}(s)}{e}\right],
\label{poisson3}
\end{eqnarray}
where $N_{+}$ ($N_{-}$) is the positrons (electrons) number density, 
$s$ is the arc length  along the last-open field line from the surface, 
and $-e$ is the electron charge. The accelerating electric field is given by
\begin{equation}
E_{||}(s)=-\frac{\textrm{d}\Phi_{nco}}{d s}.
\end{equation}

Unless the gap is close to the light cylinder, the GJ
 charge density (\ref{GJ}) is approximated as
\begin{equation}
\rho_{GJ}\sim-\frac{\Omega B_z}{2\pi c}.
\end{equation}
We use this expression for the GJ charge density.
\subsection{Continuity equations for particles and $\gamma$-rays}
\label{cont}
As the Poisson equation (\ref{poisson3}) indicates,  
particle's distribution is needed to obtain 
the electric structure in the gap. If charged particles move along the 
magnetic field lines with the speed $\sim c$, 
the steady continuity equation without any particle sources yields
\begin{equation}
\frac{N_{\pm}(s)}{B(s)}=\textrm{ constant along magnetic field line}.
\end{equation}
However, $\gamma$-rays radiated 
by particles may convert into electron-positron pairs. In the 
outer-magnetosphere, $X\gamma
\to e^{-}e^{+}$ process has a particular 
importance with $X$-ray illumination from the neutron star surface or
the magnetosphere. 
Taking account of the pair-creation,  
 we  write down the continuity equation into the form,
\begin{equation}
\pm B\frac{d}{ds}\left(\frac{N_{\pm}}{B}\right)
=\frac{1}{c\cos\Psi}\int_0^{\infty}d\epsilon_{\gamma}
[\eta_{p+}G_++\eta_{p-}G_{-}],
\label{cont1}
 \end{equation}
where  we assume that the electric field is positive in the gap for the 
definite of sign of polarity so that $e^{+}$ (or $e^{-}$) may be 
accelerated  outward (or inward);   
$G_{p+}(s,\epsilon_{\gamma})$ (or $G_{p-}$) denotes the 
distribution function of outwardly (or inwardly) propagating $\gamma$-rays;  
$\epsilon_{\gamma}$ is the photon energy normalized by $m_ec^2$;  
$\Psi$ is the angle between the particle's motion and 
the meridional plane and is assumed to be 
$\Psi=\sin^{-1}(r\Omega\sin\theta/c)$, 
where $\theta$ is the colatitude at the point considered. 
The pair-creation rate $\eta_{p\pm}(s,\epsilon_{\gamma})$ per unit time per 
$\gamma$-ray photon with $\epsilon_{\gamma}$ is written down as 
\begin{equation}
\eta_{p\pm}(s,\epsilon_{\gamma})=(1-\mu_c)c\int_{\epsilon_{th}}^{\infty}
d\epsilon_X\frac{dN_X}{d\epsilon_{X}}\sigma_{p},
\label{etap}
\end{equation}
where $d\epsilon_X\cdot dN_X/d\epsilon_X$ 
is the $X$-ray number density between 
energies $m_ec^2\epsilon_{X}$ and $m_ec^2(\epsilon_{X}+d\epsilon_{X})$ 
, $\cos^{-1}\mu_c$ is the collision angle between 
a $X$-ray and a $\gamma$-ray, $m_ec^2\epsilon_{th}$ is the threshold energy 
for the pair-creation, and $\sigma_p$ is the pair-creation cross-section, 
which is given by 
\begin{equation}
\sigma_{p}(\epsilon_{\gamma},\epsilon_{X})=\frac{3}{16}
\sigma_{T}(1-v^2)\left[(3-v^4)\ln\frac{1+v}{1-v}-2v(2-v^2)\right],
\label{cross}
\end{equation}
\[
v(\epsilon_{\gamma},\epsilon_{X})=\sqrt{1-\frac{2}{1-\mu_c}\frac{1}
{\epsilon_{\gamma}\epsilon_{X}}},
\]
where $\sigma_{T}$ is the Thomson cross section.

Let us next consider the continuity equation for $\gamma$-rays.  
The $\gamma$-rays radiated by high energy particles 
 are beamed nearly tangentially to the magnetic field. 
Although $\gamma$-rays deviate from 
the magnetic field line on which  $\gamma$-rays were emitted, the 
one-dimensional approximation does not treat this effect correctly. 
We deal with $\gamma$-ray distribution along the magnetic field 
line. Then we can write down the continuity equation into the 
form, 
\begin{equation}
\pm B(s)\frac{d}{ds}\left[\frac{G_{\pm}(s,\epsilon_{\gamma})}{B(s)}\right]
=\frac{-\eta_{p\pm}G_{\pm}(s,\epsilon_{\gamma})+\eta_cN_{\pm}(s)}
{c\cos\Psi},
\label{cont2}
\end{equation}
where the first term of the right-hand side represents 
 annihilation of $\gamma$-rays by the pair-creation, and the second term 
represents production by the curvature radiation. The 
emissivity for the curvature radiation $\eta_c$ is given by
\begin{equation}
\eta_{c}(R_c,\Gamma,\epsilon_{\gamma})=
\frac{\sqrt{3}e^2}{hR_c}\frac{\Gamma}{\epsilon_{\gamma}}
F\left(\frac{\epsilon_{\gamma}}{\epsilon_c}\right),
\end{equation} 
\begin{equation}
\epsilon_c=\frac{3}{4\pi}\frac{hc\Gamma^3}{m_ec^2R_c},
\end{equation}
and
\begin{equation}
F(x)=x\int_x^{\infty}K_{5/3}(t)dt,
\end{equation}
where $R_{c}$ is the curvature radius of the magnetic field line, $\Gamma$
 is the Lorentz factor of particles, $K_{5/3}$ is the modified Bessel 
function of  order 5/3, $h$ is the Planck constant, and $m_ec^2\epsilon_{c}$ 
gives the characteristic curvature photon energy radiated by a particle 
with  Lorentz factor $\Gamma$. The Lorentz factor in the gap is 
obtained by assuming that the particle's motion immediately saturates in the 
balance between the electric and the radiation reaction forces,
\begin{equation}
\Gamma_{sat}(R_c,E_{||})=\left(\frac{3R_c^2}{2e}E_{||}+1\right)^{1/4}. 
\label{gamma}
\end{equation}
The unity in the brackets in  (\ref{gamma}) does not 
influence $\Gamma_{sat}$ in the gap except for near the boundaries, 
where $E_{||}=0$.  

Saturation simplifies the problem significantly, but requires a condition.  
The  $\Gamma_{sat}$ is the Lorentz factor at which 
the typical accelerating time $t_{ac}$ equals the typical 
traditional damping time $t_{d}$ 
of the particles. The particles 
whose Lorentz factor is lager than $\Gamma_{sat}$ will 
be decelerated by the radiation back reaction 
because $t_{d}<t_{ac}$. On the other hand,
the particles which have 
 $\Gamma<\Gamma_{sat}$ in the gap will be accelerated because $t_{d}>t_{ac}$. 
If the two timescales are less than crossing time $t_{cr}=W_{||}/c$,
\begin{equation}
t_{cr}\gg t_{ac}\ \textrm{and} \ t_{cr}\gg t_{d},
\label{condition}
\end{equation}
 then the particles in the gap will have the saturated Lorentz 
factor given by equation (\ref{gamma}). As has been recently showed by  
Hirotani et al. (2003), in which motion of unsaturated 
particles are solved, the condition 
$t_{cr}\gg t_{ac}$ in (\ref{condition}) is not satisfied effectively 
for some pulsars and/or part of the gap. In the present application to 
the Vela-type pulsars, $\gamma$-rays producing the peak in the spectrum emerge 
from the central part of the gap, for which use of (\ref{gamma}) provides 
a good approximation. However, emission near and the outside of the 
gap requires 
a different treatment, which is discussed in \S\ref{outrad}.
\subsection{Boundary conditions}
\label{bound}
In this subsection, we give the boundary conditions to 
solve equations (\ref{poisson3}), (\ref{cont1}), and (\ref{cont2}).

The inner ($s=s_1$) and outer ($s=s_2$) boundaries are defined so 
that $E_{||}$ may vanish,
\begin{equation}
E_{||}(s_1)=E_{||}(s_2)=0 .  
\label{bound1}
\end{equation}
 We assume that $\gamma$-rays do not come into the  gap across either of the 
boundaries, that is,
\begin{equation}
G_{+}(s_1)=G_{-}(s_2)=0.
\label{bound2}
\end{equation}
Attempting to explain the activity of pulsars, 
it is generally believed that some currents circulate 
in the magnetosphere globally. Hence we permit particles 
to come into gap through the boundaries. We parameterize 
the inflows across the inner and the outer boundaries such that,
\begin{equation}
\frac{N_{+}(s_1)}{\Omega B(s_1)/2\pi c}=j_1,\ 
\frac{N_{-}(s_2)}{\Omega B(s_2)/2\pi c}=j_2.
\label{bound3}
\end{equation}
If $j_1=1$ (or $j_2=1$), particles are injected into the gap at the 
Goldreich-Julian rate at $s=s_1$ (or $s=s_2$).
The particle's continuity equations (\ref{cont1}) give the current conservation
per unit flux tube,
\begin{equation}
j_{tot}\equiv\frac{N_{+}(s)}{\Omega B(s)/2\pi c}+ 
\frac{N_{-}(s)}{\Omega B(s)/2\pi c}=\textrm{constant along }s.
\label{bound4}
\end{equation}
By defining $j_{g}$ as the amount of created current carriers per unit flux 
tube within the gap, the conservation 
low (\ref{bound4}) yields
\begin{equation}
j_1+j_2+j_g=j_{tot}.
\label{conserv}
\end{equation} 
The global conditions, which include an interaction with the pulsar wind, 
should determine the current. In our local model, therefore,  we adopt 
$(j_{tot}, j_1, j_2)$ or $(j_{g}, j_1, j_2)$ as the set of free model 
parameters.

We have the 7 boundary conditions $(\ref{bound1})\sim(\ref{conserv})$ against
 the 5 unknown functions. This is because  
$E_{||}=0$ is imposed at both boundaries and because $j_{tot}$ (or $j_{g}$) 
is externally imposed.  
In other words, we cannot impose the conditions 
$E_{||}(s_1)=0$ and $E_{||}(s_2)=0$ for arbitrary given boundaries 
$s_1$ and $s_2$.
 By moving the  boundaries step by step iteratively, 
we seek for the boundaries that satisfy the required conditions.  
\subsection{$X$-ray field in the magnetosphere}
\label{Xray}
We need to know a soft photon field to calculate 
the pair-creation rate $\eta_{p\pm}$. 
For GeV $\gamma$-ray photons,  which dominate in the curvature 
photons radiated 
in the gap, $\sigma_{p}$ is maximized at the soft $X$-ray band. 
The observed $X$-rays are dominated 
by the surface blackbody radiation for the Vela-like pulsars, and 
by the non-thermal radiation for the Crab-like pulsars. 
In this paper, because we 
apply our model to the Vela-like pulsars (\S\ref{appli}), 
we take only the surface blackbody radiation as the soft photon field.

For the blackbody case, the photon number density between  
energy $m_ec^2\epsilon_X$ and $m_ec^2(\epsilon_X+
d\epsilon_X)$ is given by the Planck law,
\begin{equation}
\frac{dN_X}{d\epsilon_X}=\frac{1}{4\pi}\left(\frac{2\pi m_ec^2}{ch}\right)^3
\left(\frac{A_s}{4\pi r^2}\right)
\frac{\epsilon_X^2}{\exp(m_ec^2\epsilon_X/kT_s)-1},   
\label{soft}
\end{equation}
where $r$ is the radial distance from the centre of the star,
 $A_s$ is the observed emitting area and $kT_s$ 
refers to the surface temperature. We calculate the pair-creation rate 
(\ref{etap}) by using (\ref{soft}) at each point in the gap, where the 
collision angle $\mu_c$ is given by 
\begin{equation}
 \mu_c=\cos\phi_{\gamma X}\sin\theta,
\end{equation} 
with $\phi_{\gamma X}=\sin^{-1}(r\sin\theta\Omega/c)$.
\section{Curvature Radiation from the Outside of the Gap}
\label{outrad}
In the HS model, a self-consistent determination of the
 accelerating electric field is successful in reproducing the $\gamma$-ray 
spectrum in the GeV band, at which the pulsed radiation peaks 
in $\nu F_{\nu}$ diagram. However, the calculated spectrum deviates from 
 the observed that in lower energy bands. This problem is 
solved by treating properly how particle energy evolve near and beyond
 the gap boundaries.

Because the saturated Lorentz factor $\Gamma_{sat}$ given by (\ref{gamma}) 
is proportional to 
$E_{||}^{1/4}$, $\Gamma_{sat}$ does not change much in the gap, except for 
around boundaries. In such case, the curvature 
spectrum produced in the gap is similar to the spectrum that is emitted by 
mono-energetic particles, and follows $E^{-p}_{\gamma}$ with 
the photon index  $p\sim 2/3$ below the cut-off energy  $E_c$. 
However, observations indicate much soft  slopes; $E_{\gamma}^{-1.3}
-E_{\gamma}^{-1.6}$ in MeV band.
    
If one uses the saturated Lorentz factor given by equation (\ref{gamma}), 
$\Gamma_{sat}$ decreases near 
the inner and outer boundaries because $E_{||}$ decreases.  It is notable that
 $\Gamma_{sat}$ is the Lorentz factor at which $t_{ac}=t_{d}$. 
The time scales $t_{ac}$ and $t_{d}$ become 
longer near the boundaries as compared with the crossing time $t_{cr}$. 
In fact, 
the ratio of $t_{cr}$ to $t_{d}$,
\begin{eqnarray}
\frac{t_{d}}{t_{cr}}&\sim&\frac{3}{2}\frac{m_ec^2}{e^2}\frac{R_c^2}{\Gamma^3}
\times W_{||}^{-1} \nonumber \\
&=& 4\frac{0.1\varpi_{lc}}{W_{||}}\left(\frac{\Omega}
{100\mathrm{rads^{-1}}}\right)^{-1}
\left(\frac{\Gamma}{10^7}\right)^{-3}
\left(\frac{R_c}{0.5\varpi_{lc}}\right)^2,
\end{eqnarray}
is larger than unity near the boundaries. As a result, the Lorentz factor is kept greater than  
$\Gamma_{sat}$ near and beyond the boundaries. 
These particles escape form the gap
 with a large Lorentz factor $\Gamma_{out}\sim a\ few\
 times\ 10^7$ and will radiate $\gamma$-rays even outside of the gap. 
The damping length after the escape should be comparable with 
or longer than the gap width $W_{||}$. 

On these aspects, 
we take account of the  radiation from the outside of the 
gap as well as that from inside.  
Although Hirotani et al. (2003) calculated $\Gamma_{out}$ numerically by 
solving the unsaturated motion of particles in the gap, we shall estimate  
$\Gamma_{out}$ semi-analytically; we assume that 
the particles are ejected from the gap with the Lorentz factor given by 
a saturated value at $t_{cr}=t_{d}$. On leaving the gap, 
 particles simply lose their energy by 
 radiation, because $E_{||}$ is screened out. The Lorentz factor 
of the particles ejected from the gap decreases as 
\begin{equation}
\frac{d\Gamma(s)}{ds}=-\frac{2}{3}\frac{e^2\Gamma^4(s)}{m_ec^2R_{c}^2(s)}.
\label{gameq}
\end{equation}
 The rate of decrease in $\Gamma$ per unit length 
is proportional to $\Gamma^4(s)$ if $R_{c}$=const. 
Then, the spectrum of $\gamma$-rays from the outside of the gap is the same as 
the spectrum made by particles with an energy distribution of
 \begin{equation}
\frac{dN_e(\Gamma)}{d\Gamma}\propto\Gamma^{-4}. 
\label{plaw}
\end{equation}
In short, we obtain a power-low $\gamma$-ray spectrum by superposing 
the curvature radiation at different positions; therefore, 
we need not assume a power energy distribution for particles to explain 
the observed power spectrum.

The curvature photons emitted by the particles obeying equation (\ref{plaw}) 
exhibits a soft, power-low spectrum with a photon index around 
$p= 5/3$ for $E_{\gamma}<E_{out}$ 
where $E_{out}=(3h/4\pi)(\Gamma_{out}^3c/R_c)$ is the critical energy 
for the highest-energy particles. The soft, power-low spectrum will extend 
down to the energy $E_{min}=(3h/4\pi)(\Gamma_{min}^3c/R_c)$, where 
$\Gamma_{min}$ is the Lorentz factor below which the $\gamma$-rays 
are beamed out from the line of the sight due to the magnetic curvature, 
for instance. Below $E_{min}$, the curvature spectrum becomes hard with a 
photon index around $p=2/3$. If we estimate $E_{min}$ with the particles energy when they escape across the light cylinder, we obtain $E_{min}\sim80$ MeV for 
the Vela pulsar. 
For example, the outwardly moving particles escape from the light cylinder 
with $\Gamma\sim 8\cdot10^6$ (for the Vela 
parameters). Thus, we expect that the calculated spectrum 
is improved between $E_{min}$  and $E_{out}$.

The $\gamma$-rays radiated towards the star may be converted into $e^{\pm}$ 
pairs by $\gamma B\to e^{\pm}B$ process and cause the pair-creation cascade. 
Cheng \& Zhang (1999) showed that the 
synchrotron spectrum from these cascade pairs has the photon index of 
$p\sim1.9$ in $X$-ray region. This spectral index is consistent with the pulsed
{\it RXTE} soft component for the Vela pulsar
(Harding et al. 2002).  
According to the model of 
Cheng et al. (2000),  the $\gamma$-rays escaping from the gap   
 through the {\it inner} boundary are much fainter than 
$\gamma$-rays through the {\it outer} boundary because the pair-creation 
occurs more efficiently near the inner boundary. In the present paper, 
the luminosity of the inwardly and the  outwardly  propagating 
$\gamma$-rays depend on the 
assumed currents $j_{1}$ and  $j_{2}$, respectively. In \S\ref{appli}, 
we will find $j_{1}\gg j_{2}$ to reproduce the observations.  
With this parameters, we have much more 
outwardly propagating $\gamma$-rays than inwardly.

\begin{figure}
\includegraphics[width=84mm, height=50mm]{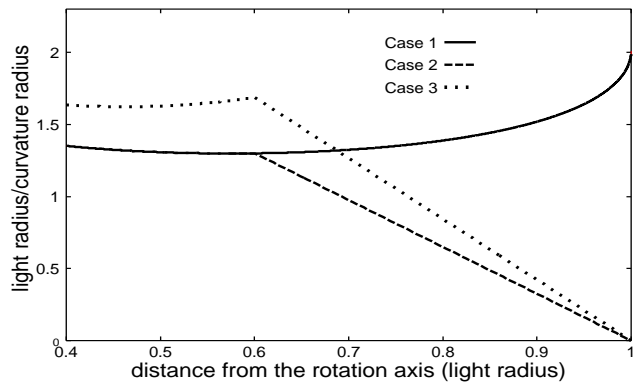}
\caption{The curvature radius of last-open field line for three cases 
when $\alpha_{in}$=45\degr. The solid, dashed-line and dotted-line 
correspond to the Case~1 (the case for pure dipole field), Case~2, and 
Case~3, respectively.}
\label{figure1}
\end{figure}

 The curvature of field lines will be  
different from that of dipole near the light cylinder.  In this paper, 
therefore, we investigate the effect of the non-dipole field on the curvature 
spectrum; this effect was not considered in Hirotani et al. (2003). 
 The many authors (e.g., Michel 1973; Contopoulos, Kazanas \& Fendt 1999) 
have attempted to set the structure of the field lines, 
but it is yet far from an enough understanding of the actual 
structure of the magnetic field (e.g., non-aligned
 rotator with the pulsar wind). To examine the variation of the curvature 
radius of field line near to the light cylinder, 
we calculate each curvature spectrum for the following three configurations:
\begin{enumerate}
\renewcommand{\theenumi}{Case~\arabic{enumi}:}
\item  the field line is defined by a  pure dipole field that is 
tangent to the light cylinder (so-called the last-open field line for the 
dipole field);
\item  the field line is the same as Case~1 below $\varpi=0.6\varpi_{lc}$, 
but deviates beyond $0.6\varpi_{lc}$ so that the radius of curvature 
may change as $R_{c}\propto1/(1-\varpi/\varpi_{lc})$;
\item Below $0.6\varpi_{lc}$, the field line is defined by a pure 
dipole field that is tangent to the cylinder of radius $0.8\varpi_{lc}$.  
Beyond $0.6\varpi_{lc}$, the field line deviates from the  pure dipole, and  
the radius of curvature of the field changes as 
$R_c\propto1/(1-\varpi/\varpi_{lc})$.
\end{enumerate}
Case~2 and Case~3 simulate the effect of the increase of open magnetic fluxes. 
In Case~3, we take account of a possibility that the last-open 
field line changes inward dipole field line as compared with 
that of  Case~1 by the influence of the pulsar wind or the electric current; 
for instance, the colatitude of the foot-point for the non-inclined  
pure dipole field line on the stellar surface is
 $\theta_f\sim (R_*/\varpi_{lc})^{1/2}$ for Case~1 (for Case~2) and 
 $\theta_f\sim (R_*/0.8\varpi_{lc})^{1/2}$ for Case~3, where $R_{*}$ is 
the star's radius. For the axisymmetric force-free magnetosphere,
 it is indicated that the field lines around the last-open field line deviate  
from dipole to radial configuration from  
$\varpi\sim 0.6\varpi_{lc}$, and 
have a point where $R_c=\infty$ near the light cylinder 
(e.g., fig.~3 in Contopoulos et al. 1999). These effects are simulated 
in Case~2 and Case~3. Fig.~\ref{figure1} shows the curvature field for 
Case~1 (solid-line), Case~2 (dashed-line), and Case~3 (dotted-line) 
between $\varpi=0.4\varpi_{lc}$ and the light cylinder when the 
magnetic inclination $\alpha_{in}$ is 45\degr.
   
The  electron-positron pairs produced by pair-creation will radiate 
by synchrotron process in the outside of the gap. 
These pairs are produced with significant pitch angle $\chi$, 
and $\Gamma\sim 5\cdot10^3$. The critical photon energy of the synchrotron 
radiation around the outer-gap becomes 
\begin{eqnarray}
E_{c}&=&\frac{3eh\Gamma^2B}{4\pi m_ec}\sin\chi \nonumber \\
&=&4.4\cdot10^5\left(\frac{\Gamma}{5\cdot10^3}\right)^2
\left(\frac{B}{10^6\textrm{G}}\right)\sin\chi \   \textrm{  eV.}
\end{eqnarray}

It is, therefore, entirely fair to neglect the synchrotron component 
emitted the outside of the gap when we consider a $\gamma$-ray spectrum 
above 100 MeV. 
%%%%%%%%%%%%%%%%%%%%%%%%%%%%%%%%%%%%%%%%%%%%%%%%%%%%%%%%%
\begin{table*}
\begin{minipage}{116mm} 
\caption{The observed parameters}
 \label{parameter}
 \begin{tabular}{@{}lccccc}
  \hline
\hline
  Pulsar & $d$ & $\Omega$
        & $B_{12}$
        & $kT_s$ & $A_s$ \\
  &kpc & rad/s & $10^{12}$G & eV
        & $\mathrm{km^2}$  \\
  \hline
  Vela & 0.5 &70.6  &3.4 &150 &$\mathrm{6.6(d/0.5kpc)^2}$ \\
  B1706-44 & 1.8 (DM)/2.5 (H I) & 61.6 & 3.1 & 143
        & $\mathrm{13(d/2.5kpc)^2}$ \\
  \hline
 \end{tabular}

 \medskip
  $d$ is the distance from the 
earth to the pulsar. For PSR~B1706-44, we adopt  
the 1.8 kpc proposed by dispersion measure and 2.5 kpc by H I absorption as 
the $d$.
$B_{12}$ represents the 
magnetic field strength at the star surface in units of $10^{12}$G.
The $X$-ray data are referred  from \"{O}gelman, Finley \& Zimmerman (1993) for the Vela 
pulsar and Gotthelf et al. (2002) for PSR~B1706-44.
\end{minipage}
\end{table*}
\section{Application}
In this section we apply our model to the twin pulsars,  
PSR~B0833-45 (Vela) and B1706-44, whose observed 
parameters are listed in Table \ref{parameter}.     
 The model parameters are ($j_{tot}, j_1, j_2$), the 
 magnetic inclination angle $\alpha_{in}$ between the 
axes of rotation and magnetization, and the cross section area $A_{gap}$ 
of the gap.  The model parameters $(j_{tot}, j_1, j_2)$ are chosen so as to 
reproduce the observed spectral properties. We adjust $A_{gap}$, which should  
not exceed $(\varpi_{lc})^2$, to the observed flux. For PSR~B1706-44, 
the proposed distance to the pulsar is 
about 1.8 kpc by the dispersion measure and about 
2.5 kpc by the H I absorption. The $X$-rays emission from PSR~B1706-44 
were reported for the first time by Gotthelf et al. (2002) with 
the {\it Chandra X-ray observatory}.
\label{appli}
\subsection{The electric structure}
\label{electric}
\begin{figure}
\includegraphics[width=84mm, height=50mm]{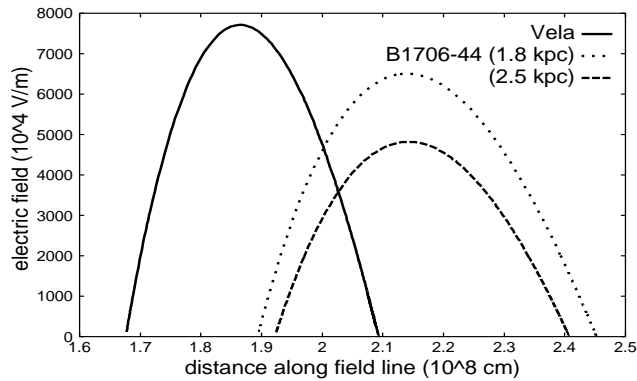}
\caption{The acceleration field for the Vela pulsar (solid-line) and 
PSR~B1706-44 (dotted-line for $d=1.8$ kpc and dashed-line for $d=2.5$ kpc). 
($j_{tot}, j_1, j_2$)=(0.201, 0.191, 0.001) and $\alpha_{in}=45\degr$}
\label{figure2}
\end{figure}

In Fig.~\ref{figure2}, we present the calculated $E_{||}$ for Case~1 
for both pulsars. The model parameters are $(j_{tot}, j_1, j_2)=
(0.201,0.191,0.001)$ and $\alpha_{in}=45\degr$. 
The abscissa refers to the arc length from the pulsar surface along 
the last-open field line.  

We see that the gap width $W_{||}$ is shorter 
than $\varpi_{lc}$, which is $4.25\times10^8$cm for the Vela pulsar and 
$4.86\times10^8$cm 
for PSR~B1706-44. In the present model, $W_{||}$ is characterized by the mean 
free path for the pair-creation as Appendix \ref{appendix1} shows. 
Since the gap is filled with abundant $\gamma$-rays and $X$-rays,  
the mean-free path is shorter than $\varpi_{lc}$.  
The ratio of the mean-free path for the Vela pulsar and that for  
PSR~B1706-44 becomes $\sim0.9$. As a result, we obtain the shorter 
gap width for the Vela pulsar (solid-line) than that for PSR~B1706-44.
   
As Fig.~\ref{figure2} shows, 
a greater  $E_{||}$ is obtained for the Vela pulsar than
for  PSR~B1706-44. This can be understood from the fact 
that the strength of $E_{||}$ 
increase with $W_{||}$  and  with the 
magnitude of $\rho_{GJ}$ in the gap. 
 The ratio of $\rho_{GJ}$ for the Vela pulsar and that for  
 PSR~B1706-44 is $\sim 2$. The effect of the difference in 
 $\rho_{GJ}$ overcomes that in the gap width, that is, PSR~B1706-44 has 
a greater gap width $W_{||}$ than the Vela pulsar does.

Fig.~\ref{figure2} also shows how $E_{||}$ depends on the 
assumed distance for B1706-44. The gap width for 
$d=1.8$ kpc (dotted-line) is wider than that for $d=2.5$ kpc (dashed-line) 
because the $X$-ray luminosity decreases with decreasing 
assumed distance. The decreased $X$-ray photon density results in an increased 
pair-production mean-free path, and hence increased gap width. 
Therefore, if we adopt a nearer distance to the 
pulsar, the calculated $E_{||}$ becomes large as Fig.~\ref{figure2} shows. 
This also causes a harder spectrum in the case of $d=1.8$ kpc 
than the case of $d=2.5$ kpc (Fig.~\ref{figure4}). 
 
\subsection{The $\gamma$-ray spectrum}
\label{spectrum}
\begin{figure}
\includegraphics[width=84mm, height=55mm]{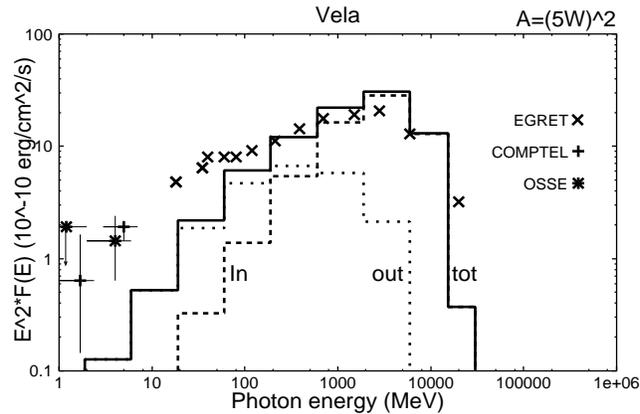}
\caption{The calculated $\gamma$-rays spectrum for the Vela pulsar. 
The total spectrum 
(solid-line) includes the gap emission (dashed-line) and 
the emission from the outside of the gap
 (dotted-line). The model parameters are $(j_{tot}, j_1, j_2)$=
(0.201, 0.191, 0.01), $\alpha_{in}=45\degr$ and $A_{gap}=(5W_{||})^2$. 
The observed points are phase average spectra which are referred from 
Kanbach et al. (1994) for EGRET and Strickman et al. (1996) 
for OSSE and COMPEL. }
\label{figure3}
\end{figure}

The dimensionless current parameters are chosen so as to reproduce the 
observation. We obtain the best-fitting parameters $(j_{tot},j_1,j_2)=
(0.201,0.191,0.001)$ for the Vela pulsar with $\alpha_{in}=45\degr$. 
 Fig.~\ref{figure3} shows the best-fitted curvature spectrum (solid-line) 
for outwardly propagating $\gamma$-rays, which 
is composed of radiation from the inside (dashed-line) and 
the outside (dotted-line) 
of the gap for Case~1. For comparison, we plot the phase averaged EGRET 
spectrum. It's peak and cut-off energies are 
in good agreement with the EGRET observations. 

As we expected in \S\ref{outrad}, the spectrum from the outside of the gap 
appears between $E_{out}\sim$ 1 GeV and $E_{min}\sim$ 80 MeV 
with the photon index $p\sim5/3$, where $E_{out}$ and $E_{min}$ correspond to 
the critical energy of curvature radiation of 
 the particle's Lorentz factor at the boundary and 
the light cylinder, respectively. Below $E_{min}$, the spectrum has 
$p\sim 2/3$. By including the emission from outside of the gap,  
we can improve the spectrum above 100 MeV.

It follows from Fig.~\ref{figure3} that  the calculated total spectrum in 
100 MeV-GeV bands becomes slightly harder than the EGRET data. 
This is because we assume that the particle's motion is saturated at 
the equilibrium Lorentz factor (\ref{gamma}), which overestimates the 
particle energy, and hence the $\gamma$-ray luminosity. A better predicted 
spectrum could be obtained if we would consider a unsaturated motion for the 
particles as Hirotani et al. (2003) did.  Nevertheless the 
outside gap emission can also be treated properly with the Lorentz factor 
solved with equation (\ref{gameq}).

\begin{figure}
\includegraphics[width=84mm, height=55mm]{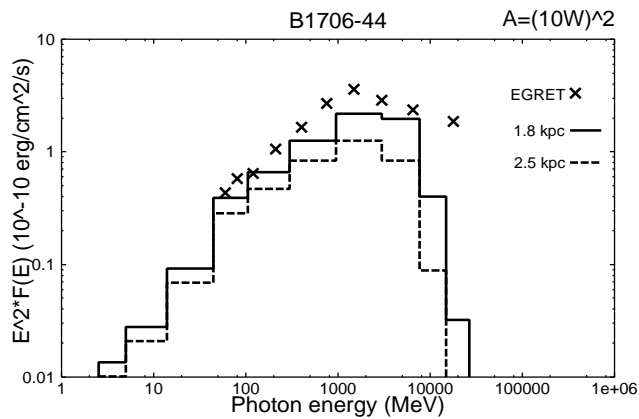}
\caption{The calculated $total$ $\gamma$-rays spectrum for PSR~B1706-44. 
The solid-line, dashed-line represent the spectra for $d=1.8$kpc and $d=2.5$ 
kpc.
The model parameters are the same as in Fig.~\ref{figure3} except for 
$A_{gap}=(10W_{||})^2$. The observed points are referred from Thompson et al. (1996). }
\label{figure4}
\end{figure}

Fig.~\ref{figure4} shows the $total$ spectrum for PSR~B1706-44. It is 
noteworthy that the same dimensionless parameters $(j_{tot}, j_1, j_2)$,
 and $\alpha_{in}$ as the Vela pulsar reproduce PSR~B1706-44 data as well.
However the observed peak energies for the two pulsars are different: 
About 3 GeV for the Vela pulsar, and about 1 GeV for PSR~B1706-44. 
 The difference is mainly due to  the 
rotation periods. As showed in 
\S\S\ref{electric}, the difference in $\rho_{GJ}$ for the two pulsars 
appears as the  difference of $E_{||}$. 
  As a result, the saturated Lorentz factor given by equation (\ref{gamma}) 
and the obtained spectral peak energy for  PSR~B1706-44 
are less than those for the Vela pulsar. Thus, we need not change 
dimensionless parameters $(j_{tot}, j_1, j_2)$. 

We fit the flux, by changing the cross section of the gap, $A_{gap}$. 
We use $A_{gap}=(5W_{||})^2\sim(0.5\varpi_{lc})^2$ 
for the Vela pulsar, and $A_{gap}=(10W_{||})^2\sim\varpi_{lc}^2$ for 
PSR~B1706-44. The value of $A_{gap}$ for B1706-44 appears to be too large, 
since the gap is located at $s \sim 
0.5\varpi_{lc}$. In fact, we have $A_{gap}/4\pi(0.5\varpi_{lc})^2\sim1/3$
 for B1706-44, which means that  
the hemisphere are almost covered with the gap.  Such 
a large $A_{gap}$ may be unrealistic. 
This flux problem will be discussed in \S\ref{dicus}.

\begin{figure}
\includegraphics[width=84mm, height=55mm]{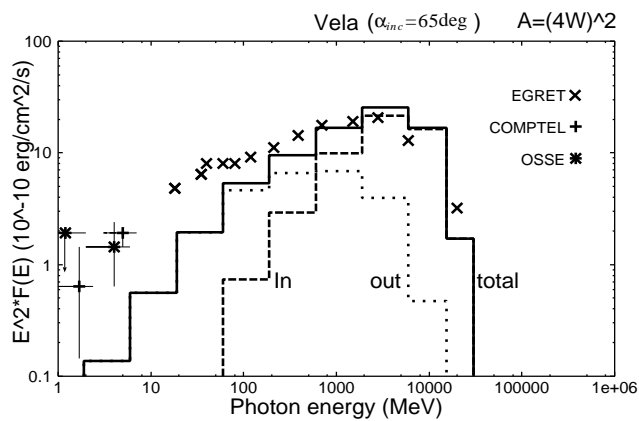}
\caption{The same as Fig.~\ref{figure3} but for $\alpha_{in}=65\degr$.
The current parameters are ($j_{tot}, j_1, j_2$)=(0.202, 0.192, 0.02) and 
$A_{gap}=(4W_{||})^2$}
\label{figure5}
\end{figure}

Let us examine how the spectrum depends on the magnetic inclination, 
$\alpha_{in}$. Romani \& Yadigaroglu (1995) estimated the inclination 
angle to be $\alpha_{in}\sim 65\degr$ 
for the Vela pulsar and $\alpha_{in}\sim 45\degr$ for PSR~B1706-44 to 
explain the 
phase difference between the radio and $\gamma$-ray pulses. We also show 
the calculated spectrum in the case $\alpha_{in}=65\degr$ for the Vela pulsar 
in Fig.~\ref{figure5}, where we adopt  
$(j_{tot}, j_{1}, j_2)=(0.202, 0.192, 0.002)$ to explain the 
observed spectral peak energy.  The strength of magnetic field in the gap
increase with $\alpha_{in}$ because the crossing point 
of the null surface and  the last-open field line comes close to 
 star. On the other hand, the gap width is shortened because  the increased 
soft photon density results in the 
decreased pair-creation mean-free path for the inwardly propagating
$\gamma$-rays.
The former effect tends to increase $E_{||}$, while the latter to decrease
 it; therefore, the two effects compensate each other. 
As a result, we obtain the observed spectral peak 
energy by using almost the same current parameters for 
$\alpha_{in}=45\degr$. On these ground, we suggest that 
almost the same currents in unit of the GJ value are running 
through the gap for the Vela pulsar and PSR~B1706-44. 

\subsection{Spectral dependence on the curvature of field line}
\label{config}
\begin{figure}
\includegraphics[width=84mm, height=55mm]{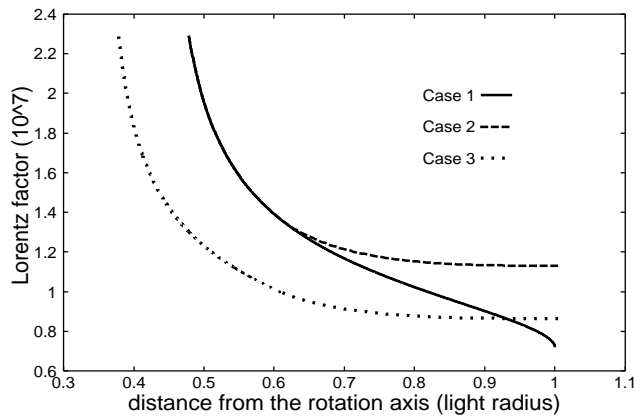}
\caption{The development of the particle's Lorentz factor in the  outside of 
the gap. The solid, dashed-line and dotted-line correspond 
to the Case~1, 2, and 3.
The abscissa is the distance from the rotation axis in unit of $\varpi_{lc}$. }
\label{figure6}
\end{figure}

Let us consider other configurations of the field line. 
Fig.~\ref{figure6} represents how 
the particle's Lorentz factor changes along the field lines with 
different field 
line configurations, where the Vela parameters are used. Three cases, 
Case~1 (solid), Case~2 (dashed) and Case~3 (dotted), are examined 
as had  been  described in \S\ref{outrad}. The 
Lorentz factors for Case~2 and Case~3 change little 
 between $\varpi=0.6\varpi_{lc}$ and $\varpi_{lc}$ because of the 
stretched magnetic field lines. 
 
\begin{figure}
\includegraphics[width=84mm, height=55mm]{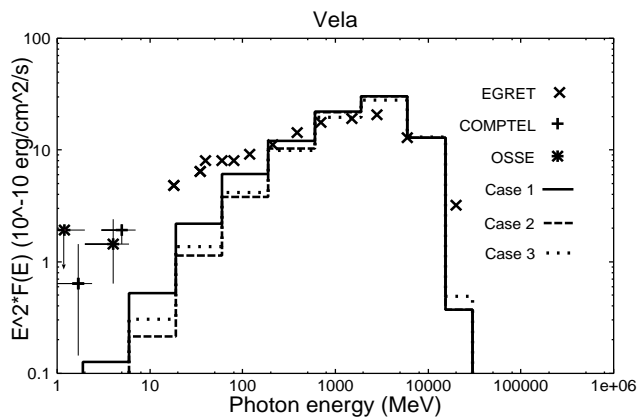}
\caption{The calculated $total$ $\gamma$-ray spectra for the  Vela pulsar. 
The lines correspond to the same case as in Fig.~\ref{figure6}. 
The cross section is chosen such that 
$A_{gap}=(5W_{||})^2$ for Case~1 
and Case~2, $(4W_{||})^2$ for Case~3.}
\label{figure7} 
\end{figure}

Fig.~\ref{figure7} shows the calculated total spectra for the three cases. 
The dependence of each configuration of the field line appears  in
the spectra around 100 MeV. On such a stretched field line 
as in Case~2 or Case~3, curvature process little contributes to the
 spectrum. Because of the less efficient curvature emission, particles escape 
across the light cylinder with a large energy as compared with a pure dipole 
field. As a result, we obtain a greater lower-cut off energy 
$E_{min}\sim 200$ MeV for Case~2 (cf. $E_{min}\sim 80$ MeV for Case~1).
   
This effect is expected to appear in the observed spectra. Therefore, 
it is possible to diagnose  at some level 
the actual field configuration in the magnetosphere from the observed 
spectra. For example, the EGRET, COMPTEL and OSSE observations indicate 
that the spectrum turns over below 100 MeV, that is, $E_{min}<100$ MeV. 
Possibly  a dipole-like field is preferable near the light cylinder 
than a stretched field, which is  expected from magnetohydrodynamics. 
 
The effect of stretched fields will be manifest in a phase-resolved 
spectrum rather than a phase-averaged one. 
 According to the double-peaked $\gamma$-ray models 
(e.g., Romani 1996; Cheng et al. 2000), 
the $\gamma$-ray photons radiated near the light cylinder 
contribute to the {\it first pulse-phase}, which leads the bridge-phase.
 Therefore, the stretching effect is expected to appear in 
the spectrum of the first pulse-phase. A diagnosis of the field line
 curvature is, therefore, carried out simultaneously with a 
position-pulse-phase 
mapping, which is necessary when we model the gap in a three-dimensional space.
 Thus, one can examine the field line geometry by using the fact that the  
curvature spectrum around 100 MeV depends on the field line geometry.
 On the other hand,  in the  phase-averaged spectrum, the radiation on 
stretching fields may be buried in the curvature radiation of 
low energy particles near the upper surface in the gap, 
where $E_{||}$ is small owing to the screening effect.

Even if a  possible variation of the field line curvature is taken 
into account, 
the discrepancy in the COMPTEL and OSSE bands cannot be resolved. It 
seems difficult to diagnose the field line with these observations. 
In general, if the curvature radius varies smoothly on the particle's path  
to the light radius,
we are able to think that the particles
escape from the light cylinder with the Lorentz factor at which 
$ct_{d}\sim\varpi_{lc}$,
\begin{equation}
\Gamma_{lc}\sim 1.17\times10^7\left(\frac{\Omega}{100\mathrm{s}}\right)^{-1/3}
\left(\frac{<R_c>}{\varpi_{lc}}\right)^{2/3},
\end{equation}
where $<R_c>$ is the averaged curvature radius of the field line along which 
the particles move. For this Lorentz factor and the radius of
 curvature, particles emit $\gamma$-rays at energies around 
\begin{equation}
E_{c}^{lc}\sim160\left(\frac{<R_c>}{\varpi_{lc}}\right)\ \mathrm{MeV},
\end{equation}
which may  appear in the curvature spectrum 
as the lower energy cut-off $E_{min}$.  If the observed break 
at $\sim10$ MeV for the Vela pulsar were caused by this escape effect, 
we would have $<R_c>\sim0.1\varpi_{lc}$, which 
should be  too short. On these ground,
 we conclude that the curvature radiation from the primary particles 
does not account for the COMPTEL and OSSE components.   

\section{discussion}
\label{dicus}
In short, we showed that  curvature flux emitted outside of the gap
 dominates that emitted within the gap around 100 MeV. With this correction, 
the model spectrum was improved above 100 MeV. We then found that the field 
line curvature near the light cylinder affects the curvature spectrum 
in several hundreds MeV. We also showed that the difference 
in spectral peak energies 
of the Vela pulsar and PSR~B1706-44 is explained by intrinsic difference 
in the magnetospheric parameters of the individual pulsars.
 We need not change the dimensionless current parameters. 

\subsection{Dominance of $j_1$ to total current}
\label{current}
\begin{figure}
\includegraphics[width=84mm, height=55mm]{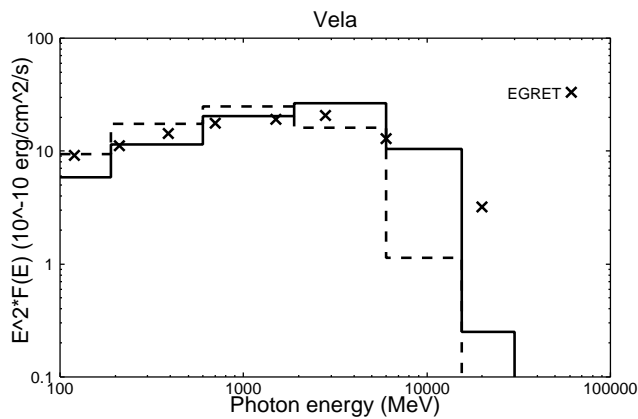}
\caption{The $\gamma$-ray spectra for the  Vela pulsar. The solid, dashed 
lines correspond to the spectra for $j_2$=0.001, 0.021, respectively, when 
$j_1=0.191$, $j_g=0.009$ and $\alpha_{in}=45\degr$. 
The cross section is chosen such that $A_{gap}=(5W_{||})^2$ for $j_2=0.001$ and $(7W_{||})^2$ for $j_2=0.021$}
\label{figure8}
\end{figure}

In \S\ref{appli}, we find that $j_{2}$ should be much smaller 
than $j_1$. In this 
subsection, we consider the case in which $j_2$ is not smaller 
than $j_1$. For a large inclination, (e.g., for $\alpha_{in}$ greater than or
 nearly equal to $45\degr$), the gap width $W_{||}$ is 
mainly determined by the pair-creation mean-free path for the 
inwardly propagating $\gamma$-rays because the collision angle 
with $X$-rays is nearly head-on for the inwardly propagating $\gamma$-rays 
and nearly  tail-on for the outwardly propagating $\gamma$-rays. 
In this case, $W_{||}$ sharply decreases with 
increase of $j_2$. Because $E_{||}$ decreases with $W_{||}$, the spectrum 
becomes soft by increasing $j_2$. Fig.~\ref{figure8} shows the spectra for  
$j_1=0.191$ and $j_g=0.009$, and $j_2=0.001$ indicated by solid-line, but  
$j_2=0.021$ by dashed-line for the Vela pulsar. A choice of $j_2=0.021$, 
which is significantly larger than the best-fitting value but smaller 
than $j_1$, does not reproduce the observed spectrum. 

Although the cases of $j_1<0.1$ can reproduce the observed 
spectral peak by appropriately choosing $j_2$, it is difficult to 
reproduce the observed flux with a cross section that is smaller 
than $(\varpi_{lc})^2$. Therefore, 
a large $j_1$ and a small $j_{2}$ are preferable. 
\subsection{Comparison with  previous models}
Romani (1996) discussed the $\gamma$-ray spectrum for the Vela pulsar  
by curvature radiation, assuming a specific energy distribution 
$dN_e/dE_{||}\propto E_{||}^{-1}$ to connect the assumed electric field, 
$E_{||}\propto 1/r$, and the number density of the particles at the 
upper surface of the gap. This relation gives the particle 
energy spectrum $dN_e/d\Gamma\propto\Gamma^{-4}$, which is identical with 
equation (\ref{plaw}), as long as the particles have the saturated 
Lorentz factor (\ref{gamma}). 
The curvature radiation of the {\it particles produced in the gap} 
dominates in the spectrum 
(see, fig.~5 in Romani 1996).

On the other hand, in the present model we obtained equation (\ref{plaw}) 
as a result of the radiative loss of the particles in the outside of the gap. 
The pair flux produced in the gap is little, 
$j_g=0.009$ for instance; therefore, the
{\it injected particles} $j_{1}$ play a crucial role in the formation of  
the curvature spectrum above 100 MeV. 

This difference in the origin of the $\gamma$-rays between the two models 
 will come from the difference in  geometrical picture of the gap,  
slab-like outer-gap model (Romani's picture) or thick gap 
model (present model).  
The slab-like gap model, the pioneering studies for which were done by  
Cheng et al. (1986a,b), assumes that the gap width along the magnetic field 
line is much larger than the trans-field thickness of the gap on the 
meridional plane. In this slab geometry, 
the number density of pairs near the upper surface of the 
gap grows exponentially in the trans-field direction by the pair-creation,  
because the  $\gamma$-rays radiated in  the gap can effectively 
across the field lines before materializing as pairs. 
In the pair-creation region, because 
$E_{||}$ is not abruptly screened,  
a copious created particles are accelerated and emit 
the observable $\gamma$-rays. 
 
In the present model, on the other hand, we assume that the 
gap width is much shorter than trans-field thickness.  In this case,
 the radiated $\gamma$-rays do not effectively across the field lines 
in the gap before materializing; therefore,  we can consider the gap with 
the one-dimensional model.  The calculated $E_{||}$  in this non-vacuum model
 decreases with increasing of pairs produced in the 
gap because of the screening effect. As a result, 
the cases of the large pair-flux production 
in the gap, the large $j_{gap}$ in (\ref{conserv}), do not reproduce 
the observed spectral peak around GeV.
As discussed in \S\S\ref{current}, therefore, we give a large injected 
particle flux into the gap through a boundary with a small $j_{gap}$ to 
explain the observations. If we consider the Romani's picture within our 
electrodynamical treatment, we should deal with a model of two- or 
three-dimensional geometry because the present one-dimensional assumption 
can not be used for such slab-like geometry. 
 
\subsection{Three dimensional electrodynamical model}
Although the difference in spectral shape of the twin pulsars, 
the Vela pulsar and
 PSR~B1706-44, is explained with the same dimensionless  current parameters, 
the difference in the flux is difficult to be understood. Because the 
two pulsars have almost the same magnetospheric parameters and the surface 
temperature, one may expect similar luminosities. If intrinsic luminosity 
were the same,  difference in flux would be attribute to the distance so
 that the expected ratio of flux may be  between 13 and 25. 
However, the observed value is about 5. 
Zhang and Cheng (1997) estimated the transverse thickness of 
the outer-gap 
by considering a condition for $\gamma$-rays to materialize marginally 
in photon-photon collisions, which may explain difference in $A_{gap}$ between 
the two pulsars. They obtained
\begin{equation}
f\sim5.5P^{26/21}B^{-4/7}_{12}  
\end{equation}
where $f$ is the transverse thickness in unit of the light radius, $P$ is the 
rotational period. With this weak dependence on the magnetospheric parameters, 
one cannot expect difference in $A_{gap}$ between the two pulsars.  One  
possibility is difference in viewing angle. For this point, we cannot 
proceeds without extending the present model to a three-dimensional one. 
We need to solve the trans-field structure of the electric field and the 
current distribution.   

There is another limitation of the present one-dimensional approach.
To get a reasonable spectrum, the dimensionless current is should be less 
than or nearly equal to  0.2 
in units of the GJ value. Within the frame work of the one-dimensional model,
this value cannot be enlarged because the gap is quenched by large currents.
This fact limits the total luminosity seriously. On the other hand,
a current of the order of the GJ value should circulate in the pulsar wind, 
and it is the  current that  runs through the gap on its way to the star
 and contributes to the spin-down.
In the outer-gap model by Cheng et al. (1986a), pairs are created 
much in the convex side of the gap, and a large current is flowing there.
To evaluate the net current in the gap and the resultant luminosity correctly,
the trans-field structure must be taken into account.

Finally, we note that the present model can only be compared 
with the observed {\it phase-averaged} spectra. 
On the other hand, the {\it phase-resolved} spectra have 
been discussed 
with three dimensional geometry such as Romani (1996) for the Vela pulsar and 
Cheng et al. (2000) for the Crab pulsar. We could  discuss 
this issue from the electrodynamical point of view if we would 
extend the present approach into a three-dimensional model
%%%%%%%%%%%%%%%%%%%%%%%%%%%%%%%%%%%%%%%%%%%%%%%%%%%%%%%%%%%%%%%%

%%%%%%%%%%%%%%%%%%%%%%%%%%%%%%%%%%%%%%%%%%%%%%%%%%%%%%%%%%%%%%%%%

\appendix
\section{The Electric Field Screening and the Gap Width}
\label{appendix1}
We give the gap width $W_{||}$ by 
 solving the continuity equations (\ref{cont1}), (\ref{cont2}) with
 some approximations.

We integrate the Poisson equation (\ref{poisson1}) along the field line  
from  $s_1$ to $s_2$ to obtain
\begin{equation}
\int_{s_1}^{s_2}[\rho(s)-\rho_{GJ}(s)]ds=0,
\label{screen}
\end{equation}
namely, the total amount of the effective charge in the gap equals zero. 
For the case of the vacuum gap ($\rho=0$), the null point, $\rho_{GJ}=0$, 
is located between $s_1$ and $s_2$.  For a given $s_1$, one can find $s_2$ 
so that (\ref{screen}) may be satisfied. The gap width is arbitrary.
 
 For a non-vacuum gap, $W_{||}$ is characterized by pair-creation mean-free 
path. Since the gap width 
is shorter than $\varpi_{lc}$ (\S\ref{appli}), we assume the 
physical quantities in the gap to be  constant value except for the 
particle's number densities ($N_{\pm}$) and the distribution function 
($G_{\pm}$) for $\gamma$-rays. Then equations (\ref{cont1}) and 
(\ref{cont2}) are combined to give
\begin{equation}
\frac{d^2N_{\pm}(s)}{ds^2}=\pm[\alpha_{+}N_{+}(s)-\alpha_{-}N_{-}(s)],
\end{equation}
 where 
\begin{equation}
\alpha_{\pm}=\frac{1}{c^2}\int_0^{\infty}d\epsilon_{\gamma}
\eta_{p\pm}(\epsilon_{\gamma})\eta_c(\epsilon_{\gamma}).
\end{equation}
With this solutions for $N_{\pm}(s)$ and the boundary conditions 
 $(\ref{bound3}), (\ref{bound4})$, we obtain the following representation 
for the gap:
\begin{eqnarray}
&&\frac{\alpha_{+}\alpha_{-}j_0W_{||}}{\sqrt{\alpha_++\alpha_-}}+
\frac{\alpha_+\left(j_2-\frac{\alpha_+j_0}{\alpha_++\alpha_-}\right)+
\alpha_-\left(j_1-\frac{\alpha_-j_0}{\alpha_++\alpha_-}\right)}
{\sinh(\sqrt{\alpha_++\alpha_-}W_{||})} \nonumber \\
&&+\frac{\alpha_+\left(j_1-\frac{\alpha_-j_0}{\alpha_++\alpha_-}\right)
+\alpha_-\left(j_2-\frac{\alpha_+j_0}{\alpha_++\alpha_-}\right)}
{\tanh(\sqrt{\alpha_++\alpha_-}W_{||})}=0.
\label{gapwidth}
\end{eqnarray}
For a largely inclined pulsar (e.g. for $\alpha_{in}$ greater than or
 nearly equal to $45\degr$), 
inwardly propagating $\gamma$-rays collide almost head-on 
with the surface blackbody 
$X$-rays. In contrast, outwardly propagating $\gamma$-rays 
collide almost tail-on. 
Due to this difference, we obtain
\begin{equation}
W_{||}\sim\frac{c}{(\int\eta_{p-}\eta_c\epsilon_{\gamma})^{1/2}}\cosh^{-1}
\left(\frac{j_2+j_{g}}{j_2}\right).
\label{gapwidth1}
\end{equation}
This equation indicates that $W_{||}$ for a large inclination 
 is characterized by the pair-creation mean-free 
path of inwardly propagating $\gamma$-rays unless $j_{2}=0$. 

By giving ($j_{tot}, j_1, j_2$), the position of the gap in the pulsar 
magnetosphere is determined by the set of $s_1$ and $s_2$
 that satisfies  relation (\ref{gapwidth}) and 
 condition (\ref{screen}). Thus, it does not need to have the 
null point in the gap.
\label{lastpage}
\end{document}